\documentclass[aps,preprint]{revtex4}%
\usepackage{amsfonts}
\usepackage{amsmath}
\usepackage{amssymb}
\usepackage{graphicx}
\usepackage{pdfpages}
\usepackage{xcolor}
\usepackage{caption}
\usepackage{lineno,hyperref}
\usepackage{changes}%
\hypersetup{colorlinks =true, allcolors = blue}
\providecommand{\U}[1]{\protect\rule{.1in}{.1in}}

\begin{document}
\preprint{HEP/123-qed}
\title{Physical Properties of Brane-World Black Hole Solutions via a Confining Potential}
\author{\.{I}zzet Sakall{\i}}
\affiliation{}
\author{}
\affiliation{}
\author{Sara Kanzi}
\affiliation{Physics Department, Arts and Sciences Faculty, Eastern Mediterranean
University, Famagusta, North Cyprus via Mersin 10, Turkey.}
\author{}
\affiliation{}

\begin{abstract}
In this paper, we consider two brane-world black holes whose solutions are obtained via a confining potential and study their thermodynamical properties. The modified entropies by taking account of the generalized uncertainty principle (GUP) are obtained. We also determine the scalar effective potentials in order to compute the greybody factors of zero-spin particles (bosons) for both geometries. Finally, using the sixth-order WKB approach, quasi-normal modes (QNMs), which are also known as black hole fingerprints, are derived concerning the brane-world black hole parameters. The results obtained are graphically depicted, tabulated, and interpreted.
\end{abstract}
\volumeyear{ }
\eid{ }
\date{\today}
\received{}

\maketitle
\tableofcontents

\section{Introduction
\ \ \ \ \ \ \ \ \ \ \ \ \ \ \ \ \ \ \ \ \ \ \ \ \ \ \ \ \ \ \ \ \ \ \ \ \ \ \ \ \ \ \ \ \ \ \ \ \ \ \ \ \ \ \ \ \ \ \ \ \ \ \ \ \ \ \ \ \ }%

According to one of the proposed universe models, our perceptible universe is a $3+1$-dimensional ($4D$) brane (the so-called extended dimension or hypersurface), which is submerged in a higher dimensional (i.e., $5D$) anti-de Sitter ($AdS_5$) spacetime, in which the only force spreads through the infinite bulk space, is gravity while other forces are constricted to the brane. In fact, the need for extra special dimensions originates from the fundamental aspects in the string theory (STH). Propounding of the extra dimensions provide the probability of the Planck scale being on an electroweak scale which relieve the problem of gauge hierarchy. It was shown that these models suffer from degradation of unitarity in about three-quarters of the low-impact Planck scale \cite{Antoniadis:2011bi}. STH is the biggest candidate to solve the problem of the unitarity. It was also found that, for the string scale it is essential to become visible lower than the effective Planck scale, in which string resonances are the first mark of these models \cite{Antoniadis:2011bi,Mavromatos:2009xg}.

In 1920s, Kaluza and Klein \cite{2,3} proposed a feasible theory for the existence of extra dimensions (the fifth dimension) point out the electrodynamics as a gravity component. Later on, in 1998, Arkani-Hamed et al  \cite{3s} presented a large extra dimensions theory in which the gravity diverges into its bulk, although the matter fields stay localized. They proved that with the extra large dimensions the famous hierarchy problem can be solved. Namely, the visible Universe might consist of a $4D$-dimensional surface (the so-called \textquotedblleft brane\textquotedblright) immersed in a $4D+D_{extra}$-dimensional spacetime (the so-called \textquotedblleft
bulk\textquotedblright). There are many investigations to consider the impression of the extra dimensions on the physics of the Standard Model (SM) and fields, which are confined to the brane \cite{3ss}.
However, the gravity has an access to reach the bulk. At least one of those additional spatial dimensions might be extremely large in comparison to the Planck scale that it causes to lower the basic gravity scale to the electroweak ($\backsim$TeV) level \cite{Maartens:2010ar}. In other words, those large extra dimensions could come with a electroweak level for quantum gravity (and/or STH) leading to the numerous new observable phenomena in colliders and other fields of the particle physics \cite{4}. The most prominent brane-world model content is  Randall-Sundrum model which was proposed in 1999 \cite{S4,S5}. They presented a different high-dimensional scenario that provides an alternative approach to constructing the hierarchy, which appear between the weak and the fundamental scales of gravity. There have been considerable investigations in the recent years about the higher-dimensional gravity where the classical $4d$ spacetime of general
relativity (GR) is recovered from the effective theories \cite{S6,N6,N7,N8}, which also admit exact solutions \cite{S7,S8}. In Ref. \cite{S9}, the construction of the Einstein field equations on a $4D$-brane planted
in a higher dimensional bulk where the matter fields are constricted to the brane in accordance with the confining potential and the associated spherically symmetric vacuum
(static) black hole solutions were investigated (see also \cite{S10,S11}).

In this paper, we consider two brane world black hole solutions having confining potentials \cite{S10,S9} in which the equation of the vacuum fields are modified by a geometrical quantity $Q_{\mu\nu}$ on the brane which may play the role of the dark energy \cite{S10}. The first solution is a good toy model to represent the galaxy's rotation curves independent of the presence of dark matter and modified Newtonian dynamics theory \cite{10SS}.  The second solution explains a black hole in the space of asymptotically de Sitter (dS). In this context, the reader is also refered to  Refs. \cite{11SS} and \cite{12SS} which do not consider the mirror symmetry that leads to a geometrical interpretation of the dark energy as bend
in the Universe given by the exterior curvature that has no mechanism needed for the confinement of matter on the brane.

In recent years, black hole thermodynamics has been repeatedly applied to test the predictions of quantum gravity candidate theories \cite{5,6,7,8}. The black hole thermodynamics in the brane-world
scenario by considering scalar and axial gravitational perturbations was studied
in Ref. \cite{9}. Besides, inspired from the inflation brane world cosmology, the thermodynamics of a black hole solution in two dimensional dilaton gravity with an arc-tangent potential background was analyzed \cite{10}. The entropy of the Reissner--N\"{o}rdstrom
black hole is regarded within the state of a brane world \cite{11}. 
However, the thermodynamics of these brane world black hole solutions have not been adequately considered under the influence of GUP. With this study, we want to fill this gap of the literature for some extent.

The emitted gravitational radiation from the horizon of a perturbed black hole releases information of its interior properties \cite{12} and makes also possible to investigate the quantum structure of the black holes. The outer side of the horizon, the potential
barrier has a frequency dependent black hole parameters that act as if a filter. As a result, some waves are reflected back by the barrier, while others are transported to infinity \cite{13}. However, the radiation emitted (or to be detected by the observer) at spatial infinity differs from its originating value measured near the black hole's horizon. More specifically, Hawking radiation changes from its perfect black body character until reaching to spatial infinity. This phenomenon is known as greybody factor (GF) \cite{14,15,16}, which reveals important information about black holes and can be computed in various methods \cite{17,18,19,20,21,22,23}.\\ 

On the other hand, QNMs play an essential role within theory of GR, which reveal the presence of gravitational radiation \cite{D23}. Furthermore, they depend on the black hole parameters and
independent of the initial disturbance to the black hole. Namely, QNMs have characteristic oscillation frequencies of the black hole \cite{24,25}. In astrophysics, the detection of QNMs from gravitational wave experiments will lead to the new tests of GR, as well as precise measurements of black hole parameters (like: mass, charge, and spin parameters). For instance, the QNMs related with gravitational-wave radiations from binary black hole coalitions can prepare deep insight into the remnant’s properties \cite{DaSilvaCosta:2017njq}. The QNMs represent as a complex form which the  real part mentions the frequency of oscillation, and its imaginary part associates with the decaying or growing (in the case of superradiance \cite{Aliev:2015wla}) characteristic time of the decay \cite{26,27}. One of the most promising approaches to compute the QNMs is the WKB method, which can be calculated in $3^{th}$ and $6^{th}$ orders \cite{D29,H11}. To wrap things up, when a black hole is perturbed, the proceeding behaviors could be expressed in three phases: The first level is correlated with the radiation because of the perturbations conditions. The second level is related to the damped oscillations with complex frequencies, (QNMs). And the third corresponds to the breakdown of a power law decay of waves \cite{Pedraza:2021hzw}.

In this paper, we study the QNMs of spherically symmetric and static (and vacuum)  black hole solutions with a confining potential in the brane-world scenario \cite{Heydar-Fard:2007ahl}. We consider two particular black hole solutions that have constant curvature bulk: (1) Schwarzschild-de Sitter black hole like solution with the existence of a cosmological constant ($\Lambda$), which can be a verification for this point that an extra term in the field equations on the brane can represented as a positive ($\Lambda$) and can be used for explaining the accelerating expansion of the universe. (2) Solution with a suitable potential explaining the galaxy rotation curves without assuming the existence of dark matter.

This paper is organized as follows: In Section 2, we introduce the Brane-World black hole solution via a confining potential (BWBHSCP). The thermodynamical propertis o BWBHSCP including its entropy with GUP are given in Sect. 3. We then study the GFs of the scalar particles in Sect. 4. To this end, we employ the Klein-Gordon equation to derive and analyze the
massless effective potential, which is efficacious in the GFs. In Sect. 5, bosonic QNMs of the BWSBHCP are numerically
computed with the aid of WKB approximation method in $6^{th}$ order. Finally, we provide conclusions in Sect. 6.

(Throughout the paper, we shall use the metric convention of $(-,+,+,+)$ and the geometrized units: $c=G=\hslash=1$.)

\section{REVIEW OF BWBHSCP SPACETIME}

The action of BWBHSCP is given by \cite{Heydar-Fard:2007ahl}%
\begin{equation}
S=\frac{1}{2\alpha^{\ast}}\int_{\aleph}D^{n}X\sqrt{-%
\mathcal{F}%
}\left(  \Re-2\Lambda^{\left(  b\right)  }\right)  +\int_{\Sigma}d^{4}%
x\sqrt{-g}\left(
\mathcal{L}%
_{surface}+%
\mathcal{L}%
_{m}\right)  , \label{1}%
\end{equation}

which, in general, stands for a spacetime $\left(  \aleph,%
\mathcal{F}%
 _{AB}\right)  $ with boundary $\left(  \Sigma,g_{\mu\nu}\right)  $. $\Lambda^{\left(  b\right)  }$ is the bulk cosmological constant and
$\alpha^{\ast}=\frac{1}{M_{\ast}^{n-2}}$ ($M_{\ast}$ represents the basic energy scale in the bulk space and $n$ denotes the dimension number). The effective field equations for vacuum
are found to be as follows \cite{Heydar-Fard:2007ahl}%
\begin{equation}
G_{\mu\nu}=-\Lambda g_{\mu\upsilon}+Q_{\mu\upsilon}-\varepsilon_{\mu\upsilon},
\label{2}%
\end{equation}

where $Q_{\mu\upsilon}$ is a geometrical quantity, which adjusts the vacuum field equations on the brane. Due to the Weyl symmetries, $\varepsilon_{\mu\nu}$  is a symmetric and traceless tensor and it is restricted by the conservation equations: $\varepsilon_{;\upsilon}^{\mu\upsilon}=0$  (ensued from the Bianchi identities). To have a brane, which has a static spherically symmetric
metric, one can use the following ansatz:
\begin{equation}
ds^{2}=-e^{\mu\left(  r\right)  }dt^{2}+e^{\upsilon\left(  r\right)  }%
dr^{2}+r^{2}\left(  d\theta^{2}+\sin^{2}\theta d\varphi^{2}\right).
\label{3}%
\end{equation}

Substituting the metric \eqref{3} into Eq. \eqref{2} and making some straightforward computations, one gets the gravitational field equations of BWBHSCP as follows:
\begin{equation}
G_{0}^{0}=\frac{e^{-\upsilon}}{r^{2}}\left(  1-r\upsilon^{\prime}
-e^{\upsilon}\right)  , \label{9}%
\end{equation}

\begin{equation}
G_{1}^{1}=\frac{e^{-\upsilon}}{r^{2}}\left(  1+r\mu^{\prime}
-e^{\upsilon}\right)  , \label{10}%
\end{equation}

and
\begin{equation}
G_{2}^{2}=G_{3}^{3}=\frac{1}{4re^{\upsilon}}\left(  2\mu^{\prime}
-2\upsilon^{\prime}
-\mu^{\prime}
\upsilon^{\prime}
r+\mu
^{\prime2}r+2\mu^{\prime\prime}
r\right)  , \label{11}%
\end{equation}

where prime symbol indicates derivative with respect to $r$. Thus, two classes of solutions are obtained. The first solution reads 

\begin{equation}
e^{\mu\left(  r\right)  }=e^{-\upsilon\left(  r\right)  }=1-\frac{B}%
{r}-\alpha^{2}r^{2}-2\alpha\beta r-\beta^{2}, \label{4}%
\end{equation}

where $B$, $\alpha$, and $\beta$ are the integration constants. However, $\alpha$ is known as the scale of energy on the brane \cite{Heydar-Fard:2007ahl}. In short-range distances regarded with conventional stellar/astrophysical scales (like our solar system scale), metric function \eqref{4} yields a Schwarzschild-like solution. On the other hand, taking into account the small $\alpha$ values which allow us to neglect the $\alpha^2$ term, the larger distance scales associated with galaxies yield the following potential function, which describes a new galaxy model as distinct from the original Einstein's GR theory:

\begin{equation}
\phi(r)=\frac{B}{2 r}+\alpha \beta r+\frac{\beta^{2}}{2}. \label{iz1}    
\end{equation}

Without referring to dark matter or assuming any new or modified theories like the modified Newtonian dynamics \cite{Turner:1997npq}, the potential function \eqref{iz1} could be utilized to explain the galactic rotation curves. The corresponding outer ($r_{+}$) and inner ($r_{-}$) horizons of the obtained metric solutions \eqref{4} can be found as 
\begin{equation}
r_{\pm}=\frac{d^{1/3}}{6\alpha}+\frac{2}{3}\left(  \frac{\beta^{2}+3}{\alpha
d^{1/3}}\right)  -\frac{2\beta}{3\alpha}, \label{5}%
\end{equation}

where%
\begin{equation}
d=8\beta^{3}-108B\alpha+12\sqrt{3}\sqrt{-4B\alpha\beta^{3}+27B%
^{2}\alpha^{2}-4\beta^{4}+36B\alpha\beta+8\beta^{2}-4}-72\beta\text{.}
\label{6}%
\end{equation}

At the considerable distance scales with
a small $\alpha$, Eq. \eqref{6} becomes

\begin{equation}
r_{\pm}=-\frac{\left(  \beta^{2}-1\right)  \pm\sqrt{\left(  \beta
^{2}-1\right)  ^{2}-16M\alpha\beta}}{4\alpha\beta}. \label{7}%
\end{equation}

For $16M\alpha\beta>\left(  \beta^{2}-1\right)^2$
there are no horizons and $16M\alpha\beta=\left(  \beta^{2}-1\right)^2$ leads to extremal black hole solution. 

The second type of solution is obtained by
considering $\beta=0$ and $\alpha\neq0$:
\begin{equation}
e^{\mu\left(  r\right)  }=e^{-\upsilon\left(  r\right)  }=1-\frac{B}%
{r}-\alpha^{2}r^{2}. \label{8}%
\end{equation}

If we consider the GR limit and compare the above result asymptotically with the black hole solution in de-Sitter space with $B=2M$, the cosmological constant is determined as $\Lambda=3\alpha^{2}$. Besides, neglecting the
second order term of $\alpha,$ the Schwarzschild horizon $r=2M$ is recovered. Also, by assuming $M=0$, one can obtain the de-Sitter horizon: $r=\frac{1}{\alpha}$.

\section{PHYSICAL PROPERTIES of BWBHSCP}

In this section, we shall study the thermodynamic and mass features of the BWBHSCP. To this end, we first recall the statistical definition of the Hawking temperature, which is determined in terms of surface gravity $\mathcal{K}$ as follows \cite{Wald,riz}:
\begin{equation}
T_{H}=\frac{\mathcal{K}}{2\pi}=\frac{1}{4\pi}\lim_{r\rightarrow r_{+}}\frac{\partial
_{r}g_{tt}}{\sqrt{g_{tt}g_{rr}}}. \label{12}%
\end{equation}

Thus, the Hawking temperatures of first \eqref{4} and second \eqref{8} kind solutions of the BWBHSCP can be computed as follows%
\begin{equation}
T_{H\left(  1\right)  }=\frac{B-2\alpha^{2}r_{+}^{3}-2\alpha\beta
r_{+}^{2}}{4\pi r_{+}^{2}}, \label{13}%
\end{equation}

\begin{equation}
T_{H(2)}=\frac{B-2\alpha^{2}r_{+}^{3}}{4\pi r_{+}^{2}}. \label{14}%
\end{equation}

In order to define the related masses, one can consider the event horizon of each BWBHSCP solution by considering $B=2M$. Therefore, we get 
\begin{equation}
M_{\left(  1\right)  }\left(  r_{+}\right)  =-\frac{1}{2}\left(  \alpha
^{2}r_{+}^{3}+2\alpha\beta r_{+}^{2}+\left(  \beta^{2}-1\right)  r_{+}\right)
,\label{15}%
\end{equation}

which also yields the mass of the second BWBHSCP solution when $\beta=0$ is set. Namely, we have 
\begin{equation}
M_{\left(  2\right)  }\left(  r_{+}\right)  =-\frac{r_{+}}{2}\left(
\alpha^{2}r_{+}^{2}-1\right). \label{16}%
\end{equation}

\subsection{GUP MODIFIED THERMODYNAMICS OF BWBHSCP}

The common feature of all promising theories in quantum gravity is that they contain the minimum observability of the Planck length $(L_{p})$. In this context, it is useful to remind the definition of the GUP \cite{N4}:
\begin{equation}
\delta p\geq\frac{1}{\delta x}\left(  1+\eta L_{p}^{2}\delta p^{2}\right)
, \label{17}%
\end{equation}

where coefficient $\eta$  is a dimensionless physical parameter that represents different quantum gravity proposals. In fact, $\eta$ is not depend on $\delta x$ and $\delta p$ but generally depends on the expectation values of $x$ and $p$ \cite{GERG}. For the energy scales much lower than Planckian scale, we can ignore the extra term, in which the usual uncertainty principle is recovered. If we re-substitute $\delta p$ into the  $\delta p$ seen in the bracket of Eq. \eqref{17}, we get%
\begin{equation}
\delta p\geq\frac{1}{\delta x}\left[  1+\frac{\alpha L_{p}^{2}}{\delta
x^{2}}\left(  1+\eta L_{p}^{2}\delta p^{2}\right)  ^{2}\right]  .\label{18}%
\end{equation}

Neglecting the higher order terms of $L_{p}$, one can have
\begin{equation}
\delta p\geq\frac{1}{\delta x}\left(  1+\frac{\eta L_{p}^{2}}{\delta x^{2}%
}\right)  . \label{19}%
\end{equation}

Recalling the invariant dispersion relation $E$ (energy) $\sim$ $P$ (momentum), Eq. \eqref{19} can be rewritten as%
\begin{equation}
\delta E\geq\frac{1}{\delta x}\left(  1+\frac{\eta L_{p}^{2}}{\delta x^{2}%
}\right). \label{20}%
\end{equation}

Thus, the energy can be defined in a more convenient form:
\begin{equation}
E\geq\frac{1}{\delta x}+\frac{\eta L_{p}^{2}}{\delta x^{3}}+O\left(
\frac{L_{p}^{3}}{\left(  \delta x\right)  ^{4}}\right)  . \label{21}%
\end{equation}

Adding higher order terms that would elaborate  Eq. \eqref{21} would of course reveal more GUP effects. Therefore, we can expand Eq. \eqref{21} as follows
\begin{equation}
E\delta x\geq1+\frac{\eta L_{p}^{2}}{\delta x^{2}}+\frac{\eta_{1}L_{P}^{3}}{\delta x^{3}}+\frac{\eta_{2}L_{P}^{4}}{\delta x^{4}}+\frac{\eta_{3}L_{P}^{5}}{\delta x^{5}},
\label{22}%
\end{equation}

where the subscript numbers are the order of derivatives of $\eta$ with respect to its argument. Now, let us consider a particle with the length $l$ and energy $E$, which is swallowed by a black hole. In such a case, the black hole involves a minimum rise area as $\Delta A\geq4(\ln2)L_{p}^{2}E\delta x$ in which we consider
$\delta x\sim l$ and $\ln2$ represents the minimum rise of entropy. By
substituting Eq. (\ref{22}) into the minimum rise area formula, we then have
\begin{equation}
\Delta A\geq4(\ln2)L_{p}^{2}\left[  1+\frac{\eta L_{p}^{2}}{\delta x^{2}}+\frac{\eta_{1}L_{p}^{3}}{\delta x^{3}}+\frac{\eta_{2}L_{p}^{4}}{\delta x^{4}}+\frac{\eta_{3}L_{p}^{5}}{\delta x^{5}}\right].
\label{23}%
\end{equation}

To compute the microcanonical entropy of the BWBHSCP, one can simply follow the methodology given in \cite{GERG} and get
\begin{multline}
\frac{dS}{dA}\approx\frac{\Delta S_{\min}}{\Delta A_{\min}}\simeq\\
\frac{1}{4L_{p}^{2}}\left[  1-\eta L_{p}^{2}\frac{1}{\delta x^{2}}-\eta_{1}L_{p}^{3}\frac{1}{\delta x^{3}}+\left(  \eta^{2}-\eta_{2}\right)\frac{L_{p}^{4}}{\delta x^{4}}+\left(  2\eta\eta_{1}-\eta_{3}
\right)  \frac{L_{p}^{5}}{\delta x^{5}}\right]  .\label{24}%
\end{multline}

By integrating the above expression, one can find the entropy as follows:
\begin{multline}
S\approx\frac{A}{4L_{p}^{2}}-\\
\pi\eta\ln\frac{A}{L_{p}^{2}}+\frac{4\pi^{3/2}\eta_{1}
L_{p}}{\sqrt{A}}-\frac{\left(  \eta^{2}-\eta_{2}
\right)  L_{p}^{2}4\pi^{2}}{A}-\frac{16L_{p}^{3}\pi^{5/2}\left(  2\eta\eta_{1}
-\eta_{3}\right)  }{A\sqrt{A}}.\label{S25}%
\end{multline}

It is worth noting that to have Eq. \eqref{S25}, the terms with order higher than $O\left(
\frac{L_{p}^{5}}{\delta x^{5}}\right)$ are neglected. By using $A=4 \pi R^{2} \simeq 4 \pi \delta x^{2}$ where $R$ is the event horizon: $R=r_{+}$. Besides, we assume that the particle acquires  $\delta x \sim R$ \cite{m25,m26} position uncertainty as it falls into the black hole. Furthermore, in the loop quantum gravity and STH, the entropy-area relationship of black holes (for $\left.A \gg L_{P}^{2}\right)$ is given by
\begin{equation}
S=\frac{A}{4 L_{P}^{2}}+\rho \ln \frac{A}{L_{P}^{2}}+\beta \frac{L_{P}^{2}}{A}, \label{S25n}
\end{equation}

where $\beta$ and $\rho$ can take various values in the content of STH and the loop quantum gravity \cite{N4}. Based on the assumption that the entropy expression given in Eq. \eqref{S25n} is more accurate and if Eqs. \eqref{S25} and \eqref{S25n} are compared, one finds out that $\eta_{1}=\eta_{3}=0$. Thus, in Eq. \eqref{22}, all terms with even power of $\frac{1}{\delta x}$ should be terminated. Namely, only even powers of Planck length could exist in the GUP. Thus, the GUP modified BWBHSCP entropy is represented as
\begin{equation}
S\approx\frac{A}{4L_{p}^{2}}-\pi\eta\ln\frac{A}{L_{p}^{2}}-\frac{\left(
\eta^{2}-\eta_{2}\right)  L_{p}^{2}4\pi^{2}}{A}.\label{27}%
\end{equation}

Applying $A=16\pi M^{2}$ in Eq. (\ref{27}), one can define the GUP assisted entropy of the BWBHSCP of first kind as follows:
\begin{multline}
S_{\left(  1\right)  }\approx\frac{\pi}{L_{p}^{2}}\left(  \alpha^{2}r_{+}%
^{3}+2\alpha\beta r_{+}^{2}+\left(  \beta^{2}-1\right)  r_{+}\right)  ^{2}-\\
\pi\eta\ln\frac{4\pi\left(  \alpha^{2}r_{+}^{3}+2\alpha\beta r_{+}^{2}+\left(
\beta^{2}-1\right)  r_{+}\right)  ^{2}}{L_{p}^{2}}-\frac{\left(  \eta^{2}-\eta_{2}\right)  L_{p}^{2}\pi}{\left(  \alpha^{2}r_{+}^{3}+2\alpha\beta r_{+}%
^{2}+\left(  \beta^{2}-1\right)  r_{+}\right)  ^{2}}.\label{N28}%
\end{multline}

In Eq. (\ref{N28}), if the GUP parameter tends to zero ($\eta\rightarrow0$) only the first term representing the non-GUP entropy \eqref{S25n} of BWBHSCP remains. Furthermore, if $\beta\rightarrow0$, then we get the GUP modified entropy of the BWBHSCP of second kind:
\begin{equation}
S_{\left(  2\right)  }\approx\frac{\pi}{L_{p}^{2}}\left(  \alpha^{2}r_{+}%
^{3}-r_{+}\right)  ^{2}-\pi\eta\ln\frac{4\pi\left(  \alpha^{2}r_{+}^{3}%
-r_{+}\right)  ^{2}}{L_{p}^{2}}-\frac{\left(  \eta^{2}-\eta_{2}
\right)  L_{p}^{2}\pi}{\left(  \alpha^{2}r_{+}^{3}-r_{+}\right)  ^{2}%
}.\label{29}%
\end{equation}

\section{GREYBODY FACTORS OF BWBHSCP}

In this section, the GFs of scalar fields radiated from the BWBHSCP are going to be analyzed. To this end, we shall reduce the radial wave equation to a one-dimensional Schr\"{o}dinger-like wave equation in terms of the tortoise coordinate. 
The Klein-Gordon equation for perturbation of massless scalar fields around a black hole can be expressed as%

\begin{equation}
\frac{1}{\sqrt{-g}}\partial_{\mu}(\sqrt{-g}g^{\mu\nu}\partial_{\nu}\Psi)=0,
\label{30}%
\end{equation}

where $\sqrt{-g}=r^{2}\sin\theta$. If we apply the above equation to the BWBHSCP line-element \eqref{3}, we get
\begin{equation}
-\frac{1}{f(r)}\partial_{t}^{2}\Psi+\left[  \frac{2f(r)}{r}\partial_{r}+f%
\acute{}%
(r)\partial_{r}+f(r)\partial_{r}^{2}\right]  \Psi+\frac{1}{r^{2}}\left[
\cot\theta\partial_{\theta}+\partial_{\theta}^{2}+\frac{1}{\sin^{2}\theta
}\partial_{\varphi}^{2}\right]  \Psi=0. \label{31}%
\end{equation}

By taking cognizance of the symmetries of the metric, we introduce an ansatz as
\begin{equation}
\Psi=e^{-i\omega t}\frac{R(r)}{r}e^{im\varphi}Y_{lm}(\theta),\label{32}%
\end{equation}

in which $R(r)$ and $Y_{lm}$ represent the radial function and spherical harmonics, respectively, $m$ represents magnetic quantum number, and $\omega$ denotes the energy of the scalar fields. Moreover, using the tortoise coordinate $r_{\ast}=\int\frac{dr}{f(r)}$ with Eq. \eqref{32} in Eq. \eqref{31}, one can obtain
\begin{equation}
\frac{d^{2}R(r)}{dr_{\ast}^{2}}+(\omega^{2}-V_{eff})R(r)=0, \label{33}%
\end{equation}

in which the effective potential can be determined as
\begin{equation}
V_{eff}=f(r)\left[  \frac{f%
\acute{}%
(r)}{r}+\frac{\lambda}{r^{2}}\right]  , \label{34}%
\end{equation}

where a prime symbol denotes the derivative with respect to $r$ and $\lambda=l(l+1)$ (recall that $l$ stands for the orbital quantum number). For the two BWBHSCP solutions, the effective potentials become
\begin{equation}
V_{eff(1)}=\left(  1-\frac{B}{r}-\alpha^{2}r^{2}-2\alpha\beta r-\beta
^{2}\right)  \left[  \frac{B}{r^{3}}-2\alpha^{2}-\frac{2\alpha\beta}%
{r}+\frac{\lambda}{r^{2}}\right]  , \label{35}%
\end{equation}

and 
\begin{equation}
V_{eff(2)}=\left(  1-\frac{B}{r}-\alpha^{2}r^{2}\right)  \left[
\frac{B}{r^{3}}-2\alpha^{2}+\frac{\lambda}{r^{2}}\right]  . \label{36}%
\end{equation}

\begin{figure}[h]
\centering\includegraphics[width=9cm,height=10cm]{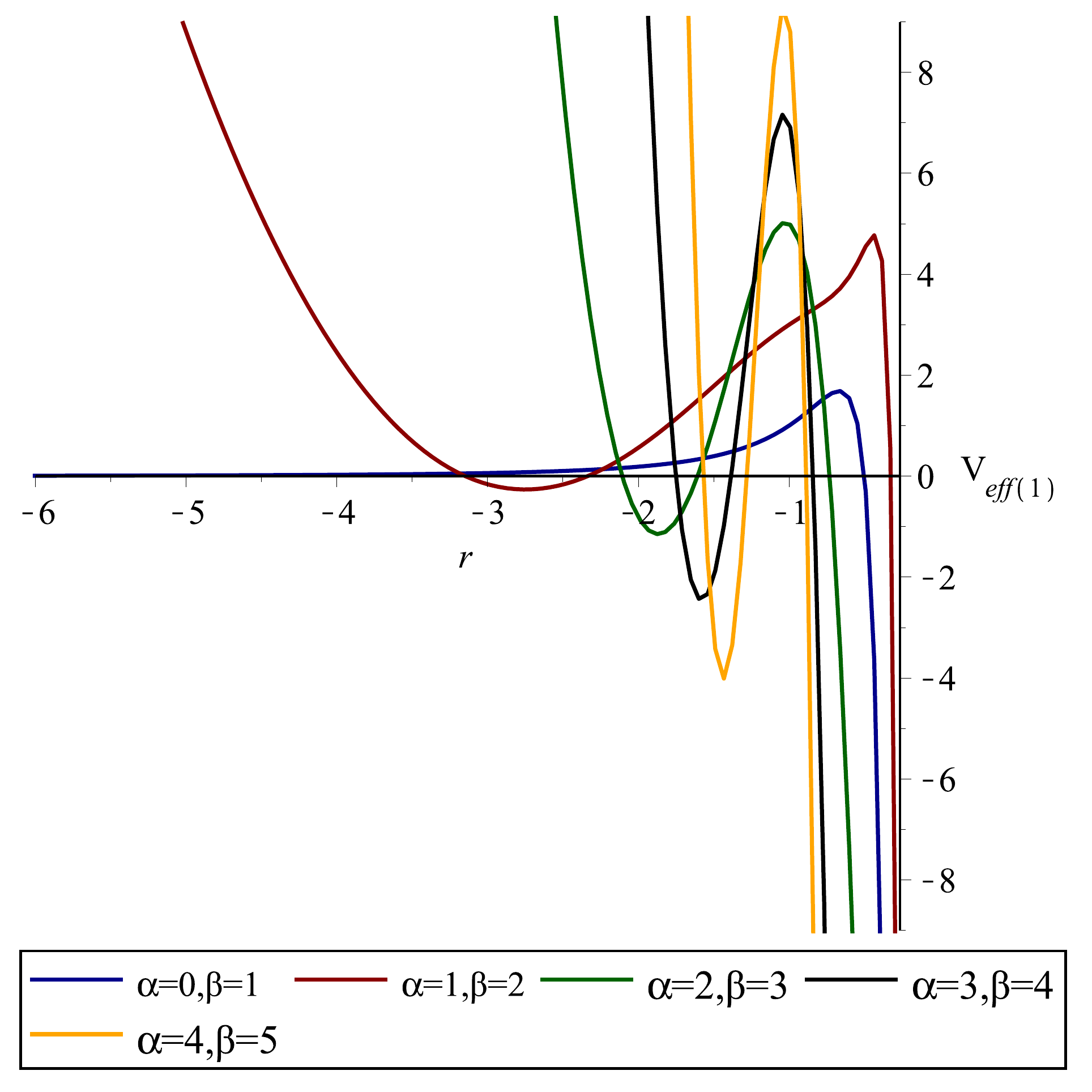}\caption{Plots of $V_{eff(1)}$ as a function of $r$ for different values of parameters $\alpha$ and $\beta$. The plots are governed by Eq. \eqref{35}, in which the parameters are chosen as  $B=1$ and $\lambda=2$.} \label{fig1}
\end{figure}

\begin{figure}[h]
\centering\includegraphics[width=9cm,height=10cm]{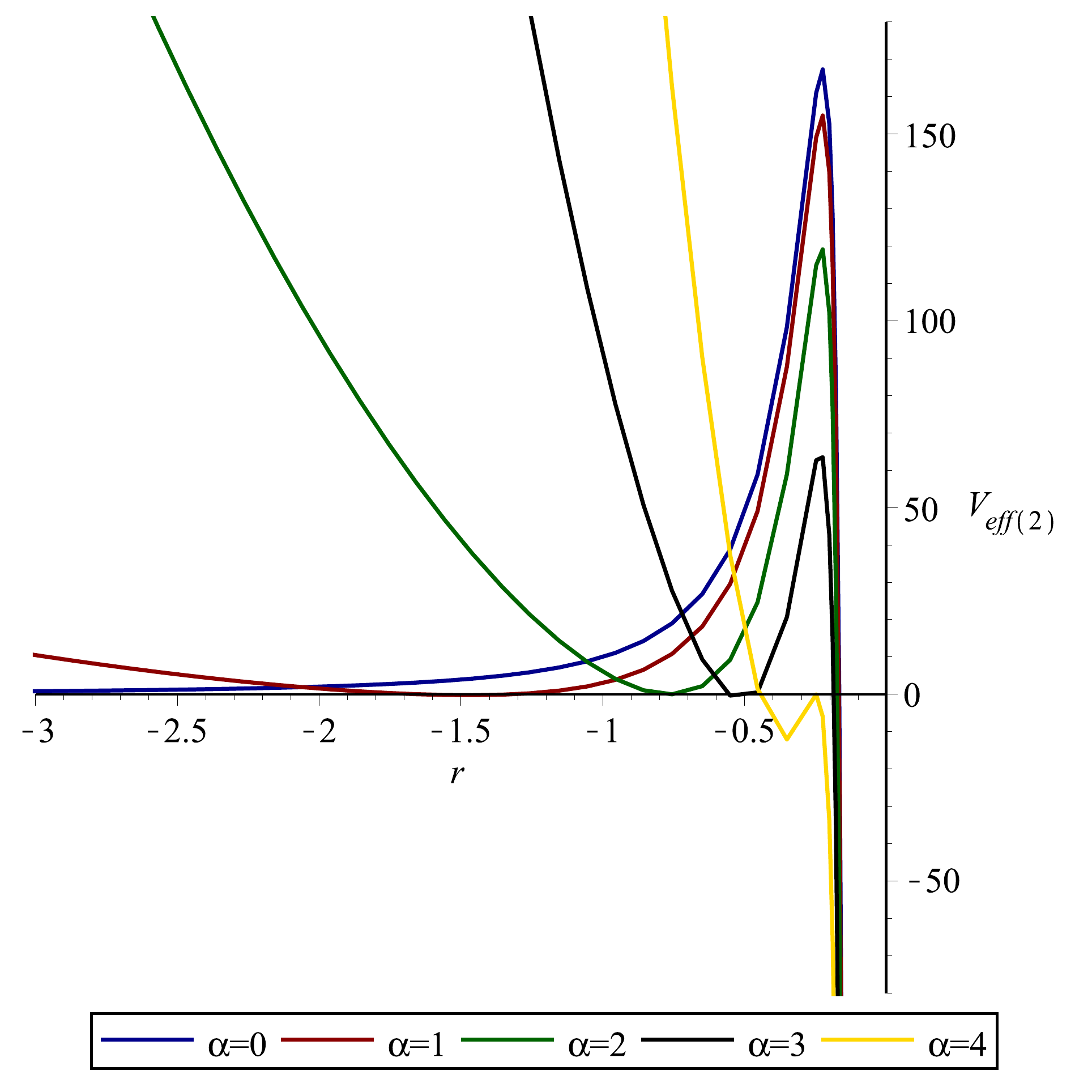}\caption{Plots of $V_{eff(2)}$ as a function of $r$ for different values of parameter $\alpha$. The plots are governed by Eq. \eqref{36}, in which the parameters are chosen as  $B=1$ and $\lambda=2$.} \label{fig2}
\end{figure}

To see the behaviors of the effective potentials, we consider Eqs. \eqref{35} and \eqref{36} and plot $V_{eff}$ versus $r$ graphs for various physical parameters of $\alpha$ and $\beta$: see Figs. \eqref{fig1} and \eqref{fig2}. 

\textcolor{blue}{One of the most important quantities in GR is GF, in which a high value of the GF implies a high probability that Hawking Radiation can reach to spatial infinity.}The GFs are known as the transmission probabilities of the wave modes generating through the relevant effective potential \cite{dd23}. Using the semi-analytical approach \cite{23}, the GFs can be computed as follows
\begin{equation}
\sigma_{l}(\omega)\geq\sec h^{2}\left\{  \int_{-\infty}^{+\infty}\wp dr_{\ast
}\right\}  , \label{37}%
\end{equation}

\textcolor{blue}{where $\sigma_{l}(\omega)$ is dimensionless GF that depends on the angular momentum quantum number $l$ and frequency of the emitted particles $\omega$}. Function $\wp$ is given by
\begin{equation}
\wp=\frac{\sqrt{(h%
\acute{}%
)^{2}+\left[  \omega^{2}-V-h^{2}\right]  ^{2}}}{2h}, \label{38}%
\end{equation}

in which $h$ is a function satisfying the following conditions 1) $h\left(r_{*}\right) > 0$ and 2) $h(-\infty)=h(\infty)=\omega$. Without loss of generality, if we simply set $h=\omega$ and replace the
tortoise coordinate with its standard definition: $r_{\ast}=\int\frac{dr}{f(r)}$, Eq. (\ref{37}) becomes%
\begin{equation}
\sigma_{l}(\omega)\geq\sec h^{2}\left\{  \frac{1}{2\omega}\int_{r_{h}%
}^{+\infty}\frac{V_{eff}}{f(r)}dr\right\}  .\label{39}%
\end{equation}

However, when using Eq. (\ref{39}), the calculations of the GFs for both BWBHSCP solutions are found to be inconclusive. The main reason of this failure is that the integral is set for the lower bound. We realized that the way to eliminate this problem is to change the method. To overcome this issue, at this stage, we employ another method which was prescribed in \cite{N5}. Thus, we first set $h=\sqrt{\omega^{2}-V_{eff}}$ in Eq. (\ref{38}). Since there are no classically forbidden zones in this approach \cite{N5}, one has%
\begin{equation}
\sigma_{l}(\omega)\geq\sec h^{2}\left\{  \frac{1}{2}\int_{-\infty}^{+\infty
}\left\vert \frac{h%
\acute{}%
}{h}\right\vert dr_{\ast}\right\}  .\label{40}%
\end{equation}

As can be seen from Figs. \eqref{fig1} and \eqref{fig2}, the effective potential peaks are related with
$h(+\infty)=h_{peak}=\sqrt{\omega^{2}-V_{peak}}$. Therefore, by setting $h(-\infty)=\omega$, Eq. \eqref{40}  thus recasts in
\begin{equation}
\sigma_{l}(\omega)\geq\sec h^{2}\left\{  \ln\left(  \frac{h_{peak}}{h_{\infty
}}\right)  \right\}  =\sec h^{2}\left\{  \ln\left(  \frac{\sqrt{\omega
^{2}-V_{peak}}}{\omega}\right)  \right\}  . \label{41}%
\end{equation}

\begin{figure}[h]
\centering\includegraphics[width=9cm,height=10cm]{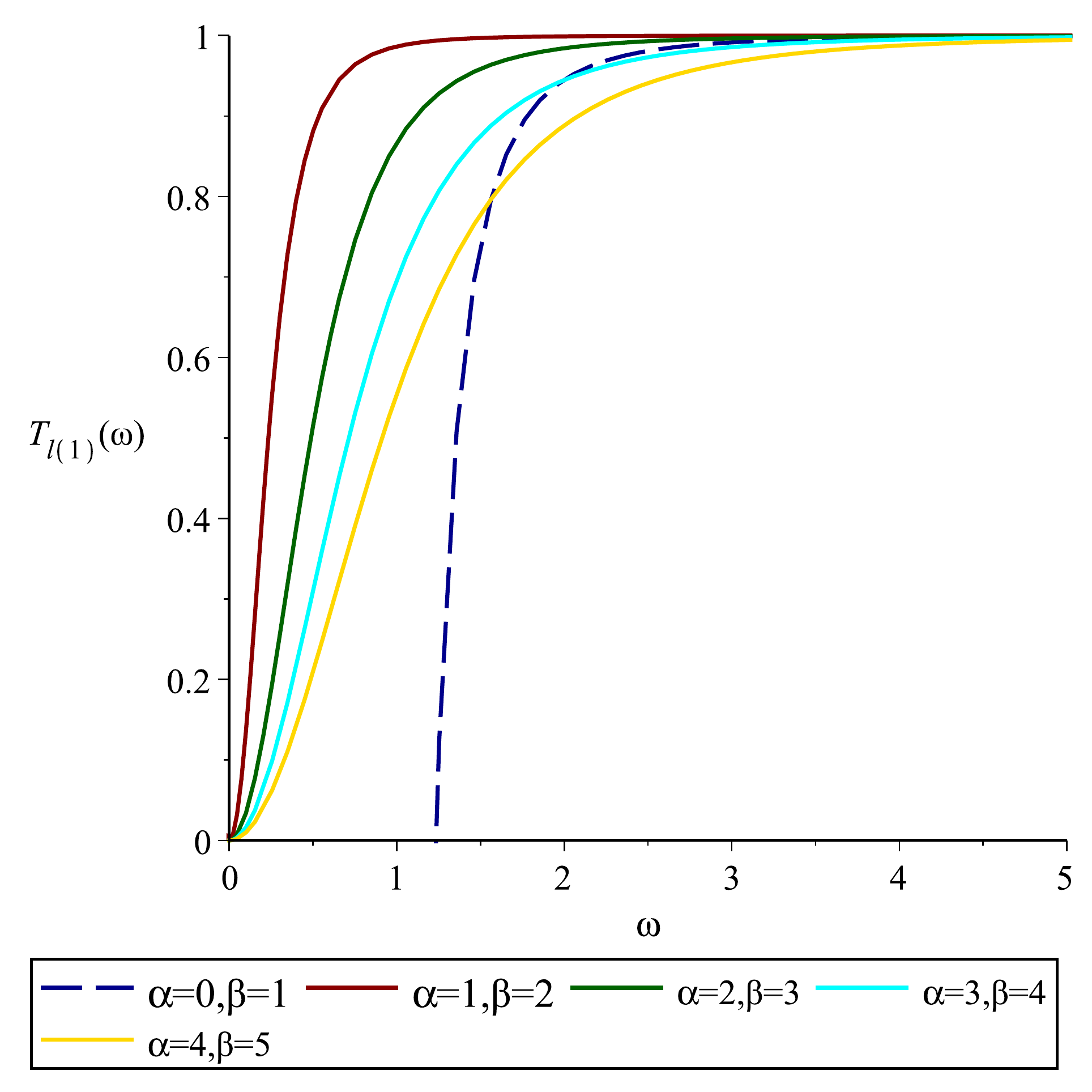}\caption{Plots of
$T_{l(1)}\left(  \omega\right)  $ versus $\omega$ for the first solution of
scalar particles. The physical parameters are chosen as; $B=1$ and
$\lambda=2$.} \label{fig3}
\end{figure}

\begin{figure}[h]
\centering\includegraphics[width=9cm,height=10cm]{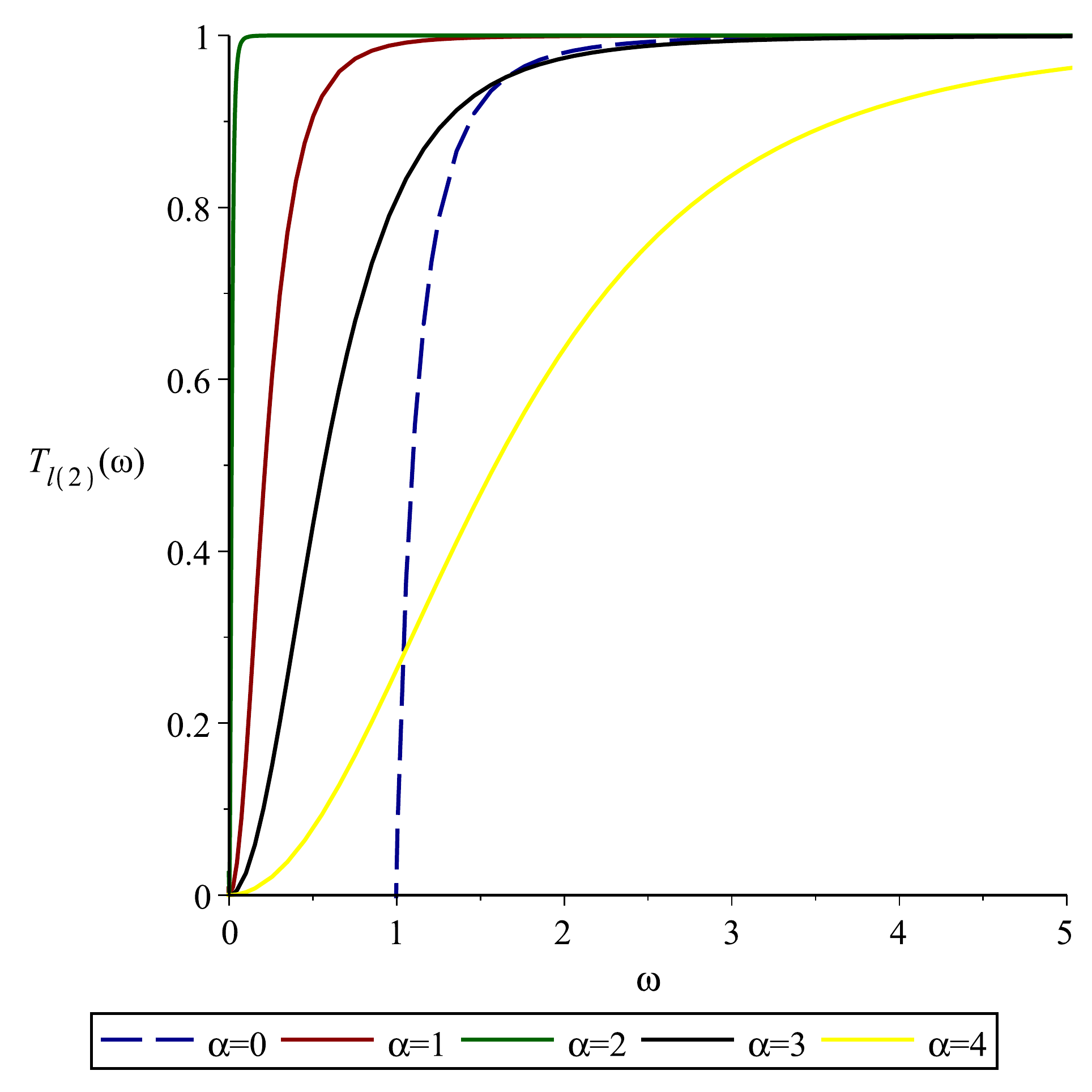}\caption{Plots of
$T_{l(2)}\left(  \omega\right)  $ versus $\omega$ for the second solution of
scalar particles. The physical parameters are chosen as; $B=1$ and
$\lambda=6$.} \label{fig4}
\end{figure}

We observe that the current bound will be pointless once $\omega^{2}<
V_{peak}$. On the other hand, one can derive the GF with the assist of transmission probability $\bigg(T_{l}(\omega)\bigg)$ via the Miller-Good
transformation method \cite{MGT}. Therefore, Eq. (\ref{41}) can be rewritten as
\begin{equation}
\sigma_{l}\left(  \omega\right)  \equiv T_{l}(\omega)\geq\frac{4\omega^{2}(\omega^{2}-V_{peak})}{(2\omega
^{2}-V_{peak})^{2}}.\label{42}%
\end{equation}
To compute $V_{peak}$, one should first define $r_{peak}$. For instance, in the first kind of solution if we let $B=1, \lambda=2$, and $\alpha=1$, $r_{peak}$ can be computed as 

\begin{equation}
r_{peak}=\frac{3\beta^{2}+3\pm\sqrt{9\beta^4-46\beta^2+73}}{8\left(\beta^{2}-1\right)}.\label{dd42}%
\end{equation}
Thus, after substituting  Eq. \eqref{dd42} into Eq. (\ref{35}), the corresponding $V_{peak}$ is obtained. One can repeat the same procedure for the second kind of solution, for different parameters as well. The GFs of BWBHSCP for some fixed parameters and various $\alpha$ and $\beta$ are depicted in \textcolor{blue}{Figs. (\ref{fig3}) and (\ref{fig4})} for the first and second kind solutions, respectively. It is worth noting that the increase in $\alpha$ and $\beta$ values decrease the GFs of the first kind BWBHSCP solution, but in the second kind BWBHSCP solution by increasing $\alpha$ the GFs grow up to $\alpha=2$ then start to decline.

\section{QNM\lowercase{s} OF BWBHSCP}

Gravitational waves from black holes possess the ringdown. The resonant frequencies that dominate the ringdown are studied by using the QNMs, which are described as waves propagating between the event horizon and spatial infinity. Namely, the QNMs of a black hole can be regarded as the solutions for the related equations of the perturbation concerned. QNMs also obey the following boundary conditions: i) pure ingoing waves at
horizon and ii) pure outgoing waves at spatial infinity.  A QNM is represented by a complex number with two types of information: the real term indicates the temporal oscillation and the imaginary term presents the exponential decay. These modes  can be detected by several large gravitational wave detectors which are still under developments \cite{sskk11}. 

In this section, for the scalar perturbation, we follow the WKB approximation method to compute the frequencies of the QNMs of the BWBHSCP. To this aim, we should simply modify the derived one-dimensional Schr\"{o}dinger- equation \eqref{33} to the following Zerilli type differential equation:%
\begin{equation}
\frac{d^{2}Z}{dr_{\ast}^{2}}+V_{Geff}Z=0,\label{43}%
\end{equation}

where $Z$ supposed to have a time-dependence $e^{i\omega t}$ and
$V_{Geff}$ is the generic effective potential. The WKB approaches can be enlarged from the $3^{rd}$ to the $6^{th}$ order \cite{25} and more \cite{dd25}. The complex frequencies of the QNMs are defined as \cite{25}:
\begin{equation}
\omega^{2}=\left[  V_{0}+\sqrt{-2V^{\prime\prime}
_{0}}\Lambda\left(  n\right)  -i\left(  n+\frac{1}{2}\right)  \sqrt{-2V^{\prime\prime}
_{0}}\left(  1+\Omega\left(  n\right)  \right)  \right]  , \label{44}%
\end{equation}
where%
\begin{equation}
\Lambda\left(  n\right)  =\frac{1}{\sqrt{-2V^{\prime\prime}
_{0}}}\left[  \frac{1}{8}\left(  \frac{V_{0}^{\left(  4\right)  }}{V_{0}^{\prime\prime}
}\right)  \left(  \frac{1}{4}+\alpha^{2}\right)  -\frac{1}{288}\left(
\frac{V_{0}^{\prime\prime\prime}
}{V_{0}^{\prime\prime}
}\right)  ^{2}\left(  4+60\alpha^{2}\right)  \right]  , \label{45}%
\end{equation}

and%
\begin{multline}
\Omega\left(  n\right)  =\frac{1}{-2V_{0}^{\prime\prime}}\left[  \frac{5}{6912}\left(  \frac{V_{0}^{\prime\prime\prime}}{V_{0}^{\prime\prime
}}\right)  ^{4}\left(  77+188\alpha^{2}\right)  -\frac{1}{384}\left(
\frac{V_{0}^{\prime\prime\prime 2}V_{0}^{\left(  4\right)  }}{V_{0}^{\prime\prime3}}\right)  \left(  51+100\alpha^{2}\right)  +\right.  \\
\left.  \frac{1}{2304}\left(  \frac{V_{0}^{\left(  4\right)  }}{V_{0}^{\prime\prime}}\right)  ^{2}\left(  67+68\alpha^{2}\right)  +\frac{1}{288}\left(
\frac{V_{0}^{\prime\prime\prime}V_{0}^{\left(  5\right)  }}{V_{0}^{\prime\prime 2}}\right)  \left(  19+28\alpha^{2}\right)  -\frac{1}{288}\left(  \frac
{V_{0}^{\left(  6\right)  }}{V_{0}^{\prime\prime
}}\right)  \left(  5+4\alpha^{2}\right)  \right]  ,\label{46}%
\end{multline}

In Eqs. (\ref{44})-(\ref{46}), the primes and superscripts ($n=4,5,6;$ for the higher order derivatives) denote the differentiation with respect to $r_{\ast
}$. The value of $r_{\ast}^{0}$ is determined and $%
\alpha=n+\frac{1}{2}.$ Considering the effective potentials \eqref{35} and \eqref{36}, the QNM frequencies are tabulated in Tables \eqref{tab1} and \eqref{tab2} for the first and second kind BWBHSCP solutions, respectively.\\

\begin{minipage}[c]{.40\textwidth}
   \centering
  \begin{tabular}{ |c|c|c|c|c| }
\hline
$l$ & $n$ & $\alpha$ & $\beta$ & $\omega_{solution(1)}$\\
\hline
1 & 0 & 0 & 0.1 & 0.286589623-0.096033341i\\
&  & 0.1 & 0.2 & 0.033656151-0.197605927i\\
&  & 0.2 & 0.3 & 0.033454334-0.280248891i\\
&  & 0.3 & 0.4 & 0.034255997+0.501632599i\\
&  & 0.4 & 0.5 & 0.152033452+0.713727366i\\
&  & 0.5 & 0.6 & 0.311345572+0.918153498i\\
&  & 0.6 & 0.7 & 0.507559367+1.112884744i\\
&  & 0.7 & 0.8 & 0.737220788+1.295439260i\\
& 1 & 0 & 0.1 & 0.258327264-0.301168481i\\
&  & 0.1 & 0.2 & 0.023903202-0.025845636i\\
&  & 0.2 & 0.3 & 0.220700417-0.414178179i\\
&  & 0.3 & 0.4 & 0.387300679-0.818680303i\\
&  & 0.4 & 0.5 & 0.512701323-1.148126858i\\
&  & 0.5 & 0.6 & 0.652993941-1.349528912i\\
&  & 0.6 & 0.7 & 0.873546922-1.356239820i\\
&  & 0.7 & 0.8 & 1.310381508-1.124252140i\\
\hline 
\end{tabular}
\captionof{table}{QNMs of scalar waves of the first kind BWBHSCP spacetime.}\label{tab1}
\end{minipage}\qquad

\begin{minipage}[c]{.40\textwidth}
   \centering
  \begin{tabular}{ |c|c|c|c| }
\hline
$l$ & $n$ & $\alpha$ & $\omega_{solution(2)}$\\
\hline
1 & 0 & 0 & 0.291114116-0.098001363i\\
&  & 0.1 & 0.029982323-0.406591230i\\
&  & 0.2 & 0.106971882-0.838143510i\\
&  & 0.3 & 0.217170609-1.295283573i\\
&  & 0.4 & 0.352035324-1.773332950i\\
&  & 0.5 & 0.504836567-2.269166503i\\
&  & 0.6 & 0.670912075-2.781643191i\\
&  & 0.7 & 0.847564070-3.310501544i\\
& 1 & 0 & 0.262211870-0.3074323461i\\
&  & 0.1 & 0.068560465-0.619918791i\\
&  & 0.2 & 0.231082654-1.302481917i\\
&  & 0.3 & 0.455973672-2.038317313i\\
&  & 0.4 & 0.725807111-2.816329955i\\
&  & 0.5 & 1.026894923-3.631446894i\\
&  & 0,6 & 1.351964324-4.482378509i\\
&  & 0.7 & 1.698488478-5.368078250i\\
\hline 
\end{tabular}
\captionof{table}{QNMs of scalar waves of the second kind BWBHSCP solution.}\label{tab2}
\end{minipage}\qquad

The results seen in Table \eqref{tab1} reveal the effect of brane-world parameters $\alpha$ and $\beta$ of the first solution on the QNMs. The oscillation frequencies (real parts) first decrease with the increasing $\alpha$ and $\beta$ parameters then start to monotonically rise, this behaviour is also observed for the damping rates (imaginary parts). The QNMs of the second solution exhibit more regular behaviours as  $\alpha$ parameter increases: both real and imaginary parts grow with increasing  $\alpha$ values.

\section{CONCLUSION}
In this paper, the GFs and QNMs of gravitational perturbations throughout the BWBHSCP geometries are investigated. To obtain the corresponding GFs and QNMs, we have perturbed the BWBHSCP backgrounds with the Klein-Gordon equation. Taking the advantage of the symmetry, we decomposed the wave function into radial and angular (with the aid of the spherical harmonics) parts. The derived radial wave equation has been used to obtain the effective potentials \eqref{34}. Using the semi-analytical approach \cite{23}, GFs of the BWBHSCP metrics have been computed and represented in Eq. \eqref{41}. The GFs of BWBHSCP for some fixed parameters and various $\alpha$ and $\beta$ are depicted in Figs. (\ref{fig1}) and (\ref{fig2}) for first \eqref{4} and second \eqref{8} kind BWBHSCP solutions, respectively. We have deduced from those figures that the increase in $\alpha$ value decreases the GFs of the both BWBHSCP solutions.  Then, we have considered the $6^{th}$ order of the WKB approach to compute the QNM frequencies. Additionally, as shown in Tables \eqref{tab1} and \eqref{tab2}, our QNM results reveal that the BWBHSCP refers the classical QNM behaviors of the Schwarzschild black hole for $\alpha \rightarrow 0$ \cite{25}. Besides, as can be observed in the tables, for some $(\alpha,\beta)$ values, the imaginary term of the QNM frequencies are negative. Therefore, our results imply that some modes are found to be unstable. Moreover, we have also studied the effect of the GUP on the BWBHSCP entropies, till the square order of Planck length.

The results of this study are promising and motivate us for
further work in this direction. Therefore, we plan to extend our study to the other type brane world
black holes, such as the rotating ones \cite{rbh1,rbh2}. 

\section*{Acknowledgements}
We are thankful to the Editor and anonymous Referee for their constructive suggestions and comments.

\end{document}